# MEMS chip-based single proof-mass triaxial fiber-optic accelerometer with ultra-low noise level

Chaoyue Liu[1], Ping Lu[1,*]

[1]National Engineering Research Center of Next Generation Internet Access System, School of Optical and Electronic Information, Huazhong University of Science and Technology, Wuhan 430074, China

**Abstract**

High-precision triaxial acceleration detection holds critical applications in seismic wave detection, geological resource exploration, and aerospace systems. Fabry-Pérot (FP) optical sensors have gained widespread adoption in these domains due to their compact footprint and immunity to electromagnetic interference. Nevertheless, conventional three-axis measurements predominantly rely on assembling multiple single-axis transducers, introducing limitations such as increased device volume and misalignment errors. In this paper, we demonstrate a MEMS-based monolithically integrated triaxial optical accelerometer that integrates a compact size with minimal noise and low crosstalk. The triaxial sensing structure employs a shared proof mass, achieving significant miniaturization compared to conventional multi-chip assembled triaxial optical accelerometers. In-plane sensing is realized through folded spring beams, while out-of-plane detection utilizes U-shaped suspension beams with widened central segments to suppress cross-axis sensitivity and enhance mechanical responsivity. Experimental results demonstrate that an operational bandwidth of 1–35 Hz, a minimum detectable acceleration of 4.12 ng/√Hz, and crosstalk below 0.023%. The compact sensor footprint measures 16 mm × 16 mm × 0.5 mm. This optical accelerometer achieves nano-g resolution in the three-axis direction, demonstrating strong potential for applications in seismic wave detection and other precision vibration monitoring fields.

## 1. Introduction

Accelerometers are widely employed to measure various acceleration signals and find extensive applications in fields such as seismic wave detection, structural health monitoring, and petroleum geological resource exploration[1-4]. The advent of MEMS (Micro-Electro-Mechanical Systems) technology has driven the miniaturization of accelerometers, enabling their use in precision applications. Currently, electrical MEMS accelerometers, primarily including capacitive[5], piezoelectric[6,7], and piezoresistive[8] types, have undergone rapid development and are well-established in practical applications. However, these sensors are susceptible to electromagnetic interference, posing challenges for their deployment in certain harsh environments. The optical accelerometer does not contain electronic components, has the advantages of anti-electromagnetic interference, chemical stability, and long transmission distance[9,10].

Low noise and compact size have long been the significant objectives in the pursuit of accelerometer development. In 2025, Wang et al. employed 3D printing technology to fabricate an optical accelerometer on the end-face of a ceramic ferrule, resulting in an extremely compact device with dimensions on the order of hundreds of micrometers and an equivalent noise acceleration of 62.45 μg/√Hz[11]. Although it possesses compact dimensions, its noise level remains relatively high.

In contrast, Jiao et al. proposed an optical accelerometer based on a plano-concave Fabry-Pérot (FP) microcavity, achieving a significantly reduced noise level of 2.5 ng/√Hz with a footprint of 10 mm × 10 mm × 0.5 mm[12]. They each possess distinct advantages in terms of size or noise performance; however, they are all single-axis sensors capable of detecting acceleration in only one direction, necessitating the assembly and integration of multiple units to detect signals along all three orthogonal directions[13-18]. And this assembly process often introduces drawbacks such as increased device size and misalignment errors.

In contrast, monolithic triaxially integrated accelerometers can overcome these limitations, which is primarily achieved through two approaches. The first approach involves fabricating three sensing structures oriented in different directions onto a single silicon substrate, thereby forming a monolithic triaxial accelerometer[19-21]. In 2025, Hong et al. fabricated an optical triaxial accelerometer featuring multiple sensing structures monolithically integrated on a single silicon chip[21]. Their device achieved a noise floor of 21.8 ng/√Hz and an orthogonal cross-axis sensitivity of 1.32%. The second approach employs a multi-degree-of-freedom triaxial accelerometer featuring a single proof mass suspended by multiple supporting beams. In contrast to the former type, this configuration achieves a more compact footprint due to the three axes sharing a common proof mass. However, this architecture presents greater challenges in ensuring triaxial sensitivity consistency and suppressing cross-axis interference. Notably, such sensors have thus far been predominantly implemented in electrical accelerometers[22,23]. Abozyd et al. first demonstrated a multi-degree-of-freedom optical accelerometer in 2022, which employed four folded beams to support three-axis motion. Although the device achieved a compact size of 4 mm × 4 mm × 1.5 mm, it exhibited relatively high noise levels, reaching 56.2 μg/√Hz[24]. Triaxially integrated accelerometers enable triaxial signal detection without the need for assembly, but the more complex structural design often results in inferior performance compared to their single-axis counterparts in terms of noise, crosstalk, and size. Consequently, the optical accelerometer urgently requires a new solution that simultaneously achieves triaxial sensing, low noise, compact size, and low crosstalk.

This paper presents a monolithic triaxial optical accelerometer based on MEMS technology, which integrates a compact size with low noise and minimal crosstalk. The sensor employs folded spring beams for in-plane (x/y-axis) sensing and a specialized U-shaped beam for out-of-plane (z-axis) sensing, with all three axes sharing the common proof mass, thereby achieving a significantly more compact footprint. Furthermore, simulation analysis verifies the U-beam's advantage in suppressing cross-axis interference. By implementing Fabry-Pérot (FP) cavities with distinct cavity lengths on the three axes, frequency-division multiplexing is realized. The proposed sensor structure measures 16 mm × 16 mm × 0.5 mm, representing a size reduction compared to previously reported optical triaxial accelerometers. The average crosstalk along the out-of-plane Z-axis is lower than 0.023%, confirming the unique advantage of the specialized U-shaped beam in suppressing mechanical crosstalk. This design approach provides a novel insight for future out-of-plane accelerometer development. The triaxial accelerometer achieves noise level of 4.12 ng/√Hz, and its extremely low self-noise provides the foundation for high-precision acceleration measurement. Therefore, achieving miniaturized triaxial measurement, low crosstalk, and extremely low self-noise enables the proposed accelerometer to have broad application prospects in fields such as seismic wave detection, oil and gas exploration, and precision aerospace systems.

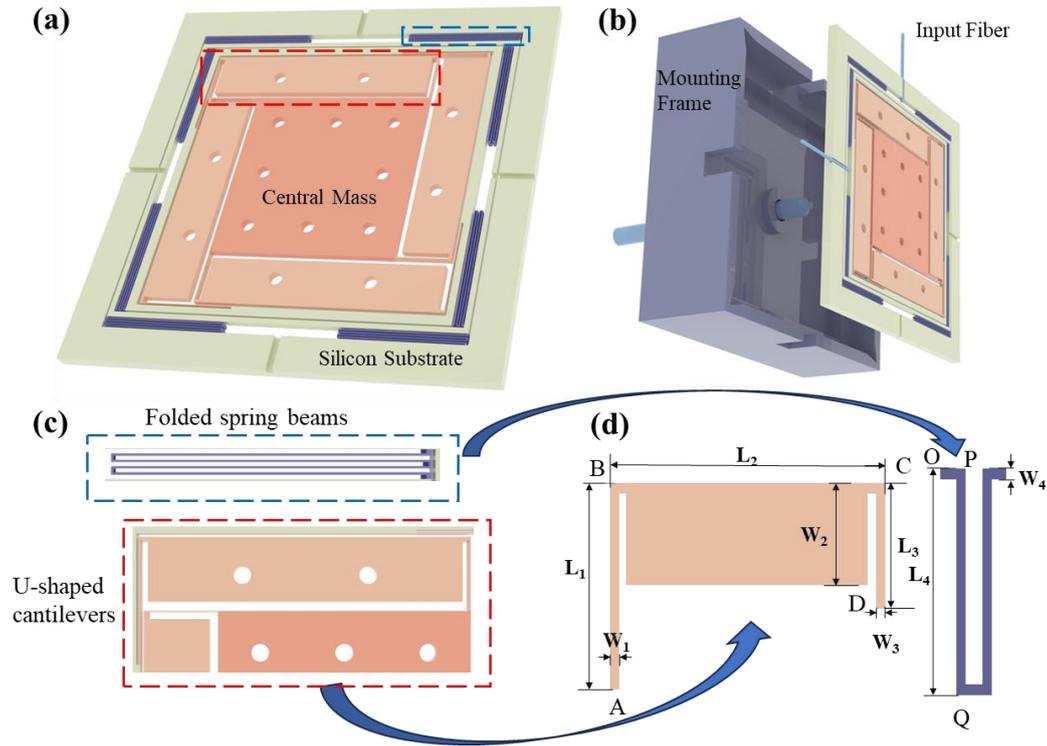

Fig. 1. The schematic of the proposed sensing structure. (a) The schematic of three-axis accelerometer with a single mass block on chip. (b) Schematic diagram of the accelerometer packaging structure. The direction within the plane is achieved by inserting and fixing two optical fibers into the grooves and forming an FP interference microcavity with the sidewall of the silicon frame. The external-direction optical fiber passes through the ceramic ferrule, and the ceramic ferrule is fixed by the through-hole of the packaging base. The fiber end face forms an FP interference microcavity with the bottom surface of the central mass block. (c) The structural enlargement drawing of the folded spring beam and the U-shaped cantilever beam. The structures in the blue box and the red box correspond respectively to those in Fig.1 (a). (d) Structure diagram of the two beams

## 2.Structure and principle
### A. Sensor Design and Simulation

A schematic diagram of the proposed triaxially integrated accelerometer is shown in Fig.1(a). The structure comprises eight folded spring beams and four U-shaped cantilever beams, all interconnecting a single central proof mass to the surrounding frame. The folded spring beams serve as the sensing elements for in-plane motion (X-axis and Y-axis), while the U-shaped cantilevers provide out-of-plane (Z-axis) sensitivity. The beam structures for each sensing direction are physically isolated by silicon frames, and the symmetric overall structure effectively constrains the central mass, contributing to reduced cross-axis interference. Appropriately increasing the beam width at the bends and connections of both the folded spring beams and the U-shaped cantilevers mitigates stress concentration, thereby enhancing the maximum allowable stress. Furthermore, multiple circular through-holes within the central proof mass and U-shaped cantilevers serve to minimize air damping during out-of-plane motion, consequently reducing the sensor's intrinsic noise. Optical fiber grooves are etched into the perimeter frame to secure optical fibers. Fabry-Pérot cavities are formed horizontally between the fiber end-face and the sidewall of the silicon frame, and vertically between the fiber end-face and the bottom surface of the central proof mass. Under acceleration, the displacement of both the proof mass and the frames modulates the respective FP cavity lengths. Demodulating the signals from these three cavities thereby obtains the triaxial acceleration data.

A critical challenge in triaxially integrated accelerometer design is minimizing cross-axis interference. The proposed out-of-plane U-shaped beam significantly suppresses in-plane to out-of-plane coupling. Applying Castigliano's second theorem, the displacement at the beam's free end can be expressed as the partial derivative of the structural strain energy with respect to the applied force. The total strain energy comprises components from axial, bending, and torsional strains. This analytical framework enables derivation of the directional stiffness coefficients for the beam structure. For the U-shaped beam illustrated in Fig. 1(d), under a horizontally applied force:

- The displacement in the X-direction at the free end is dominated by axial strain in the segments AB and CD and bending strain in the segment BC.

$$\delta_X = \int_0^{L_{AB}} \frac{F^2}{2EA_{AB}} \frac{\partial F(x)}{\partial F} dx + \int_0^{L_{BC}} \frac{(Fx)^2}{2EI_{BC}} \frac{\partial F(x)}{\partial F} dx + \int_0^{L_{CD}} \frac{F^2}{2EA_{CD}} \frac{\partial F(x)}{\partial F} dx \\ = \frac{FL_{AB}}{EA_{AB}} + \frac{FL_{BC}^3}{3EI_{BC}} + \frac{FL_{CD}}{EA_{CD}} \tag{1}$$

- The displacement in the Y-direction at the free end arises primarily from bending strain in the segments AB and CD and axial strain in the segment BC.

$$\delta_Y = \int_0^{L_{AB}} \frac{(Fx)^2}{2EI_{AB}} \frac{\partial F(x)}{\partial F} dx + \int_0^{L_{BC}} \frac{F^2}{2EA_{BC}} \frac{\partial F(x)}{\partial F} dx + \int_0^{L_{CD}} \frac{(Fx)^2}{2EI_{CD}} \frac{\partial F(x)}{\partial F} dx \\ = \frac{FL_{AB}^3}{3EI_{AB}} + \frac{FL_{BC}}{EA_{BC}} + \frac{FL_{CD}^3}{3EI_{CD}} \tag{2}$$

where $F$ is the applied force, $E$ is the Young's modulus of the material, $I_{AB}$, $I_{BC}$, and $I_{CD}$ are the area moments of inertia of segments AB, BC, and CD, respectively, and $A_{AB}$, $A_{BC}$, and $A_{CD}$ are the cross-sectional areas of segments AB, BC, and CD, respectively. Given that the four U-shaped cantilever beams are arranged in a centrosymmetric manner, and mechanically equivalent to springs connected in parallel within a harmonic oscillator model, the elastic coefficient in the horizontal direction is thus given by:

$$k_{X/Y} = \frac{2F}{\delta_X} + \frac{2F}{\delta_Y} = \left( \frac{L_{AB}}{2EW_1 h} + \frac{2L_{BC}^3}{EW_2^3 h} + \frac{L_{CD}}{2EW_3 h} \right)^{-1} + \left( \frac{2L_{AB}^3}{EW_1^3 h} + \frac{L_{BC}}{2EW_2 h} + \frac{2L_{CD}^3}{EW_3^3 h} \right)^{-1} \tag{3}$$

When a force is applied along the sensitive axis, segment AB is subjected to both a bending moment and a torque, while segments BC and CD primarily experience bending moments. Consequently, the elastic coefficient is expressed as:

$$k_Z = \frac{FL_{AB}^3}{3EI_{AB-Z}} + \frac{FL_{BC}^3}{3EI_{BC-Z}} + \frac{FL_{CD}^3}{3EI_{CD-Z}} + \frac{FL_{AB}(L_{BC} - b_3/2)^2}{GJ_{AB}} \tag{4}$$

where $J_{AB}$ denotes the torsional constant of segment AB. The structural behavior along each axis can be modeled as a spring-mass oscillator. Therefore, the mechanical sensitivity in each direction can be obtained from the elastic coefficient of the beam, which is expressed as:

$$S_m(\omega) = \frac{\Delta z(\omega)}{a(\omega)} = \frac{1}{\omega_0^2 \sqrt{(\Omega^2 - 1)^2 + (\Omega/Q)^2}} \tag{5}$$

where $\omega$ represents the angular frequency of the input acceleration signal, $\Delta z(\omega)$ is the relative

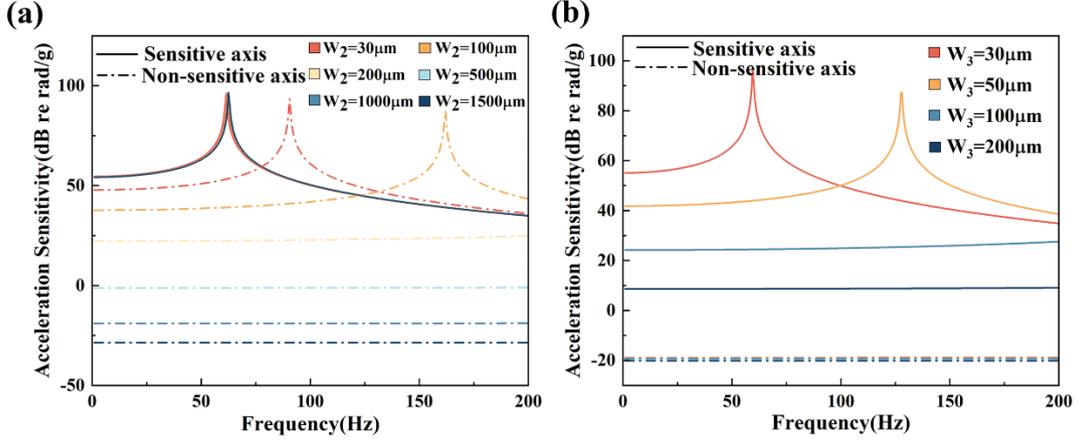

Fig.2 (a) Frequency response under varying BC segment width conditions. (b) Frequency response under varying CD segment width conditions. The solid line represents the sensitive axis, which is the sensitivity in the direction outside the plane. The dotted line represents the non-sensitive axis, which is the sensitivity in the direction within the plane.

displacement between the mass block and the fixed optical fiber, $a(\omega)$ is the amplitude of the input acceleration signal, $\omega_0$ is the resonant angular frequency of the structure, $\Omega=\omega/\omega_0$ is the ratio of the signal frequency to the resonant frequency, and $Q=(kM)^{1/2}/C$ is the quality factor of the mechanical structure, C is the damping coefficient, M is the quality of the mass block. Using the FP interferometric optical phase modulation method, the optical phase acceleration sensitivity $S_{opta}$ can be expressed as:

$$S_{opta} = \frac{4\pi n}{\lambda} S_m \qquad (6)$$

where $n$ denotes the refractive index of the optical microcavity, $\lambda$ is the center wavelength of the light source. Substitute $k$ into Equations (5) and (6) to obtain the frequency response curve. Numerical simulations were performed for varying widths of segments BC and CD, with results presented in Fig. 2. As the width of BC increases, the stiffness along the sensitive direction remains nearly constant, indicating no change in mechanical sensitivity for the Z-axis. Conversely, the stiffness in the horizontal direction increases significantly, leading to a rapid reduction in cross-axis sensitivity. By strategically increasing the width of BC, this work achieves minimized cross-axis sensitivity while preserving mechanical sensitivity along the primary sensing direction. This design shifts higher-order lateral vibration modes away from the dominant out-of-plane mode, thereby impeding energy transfer from the primary mode to higher-order modes and substantially reducing orthogonal cross-axis interference. Furthermore, widening BC effectively increases the equivalent inertial mass of the system, providing an additional enhancement to sensitivity in the sensing direction. Fig. 2(b) presents parametric analysis of CD segment width variation. The CD width significantly impacts sensitive-axis responsivity, but demonstrates limited efficacy in suppressing cross-axis sensitivity. On the other hand, the width of segment AB has no effect on the responses of both the sensitive axis and the non-sensitive axis. At the same time, the length of this three-segment structure will have an impact on the sensitivity of both the sensitive axis and the non-sensitive axis. Consequently, the length optimization should be determined primarily based on dimensional constraints and sensitive-axis performance requirements. The folded spring beam, due to its high sensitivity and low crosstalk characteristics, has been widely applied in the design of acceleration structures. In the planar structure design, this paper also adopts multi-period elastic folded springs as the beam structure. This structure mainly controls the frequency response characteristics of the planar structure by changing the length and width of the PQ section, and controls the external crosstalk by controlling the thicknes. Based on combined theoretical analysis and finite element simulation, the key structural parameters of the proposed accelerometer were finalized and are listed in Table 1.

**Table 1. Key structural parameters of the proposed sensors**

| Parameter | Symbol | Value/μm | Parameter | Symbol | Value/μm |
|---|---|---|---|---|---|
| Spring width 1 (AB) | $W_1$ | 37 | Spring length 1 (AB) | $L_1$ | 2060 |
| Spring width 2 (BC) | $W_2$ | 1800 | Spring length 2 (BC) | $L_2$ | 9200 |
| Spring width 3 (CD) | $W_3$ | 32 | Spring length 3 (CD) | $L_3$ | 3560 |
| Spring width 4 (OP) | $W_4$ | 32 | Spring length 4 (PQ) | $L_4$ | 4500 |
| Thickness | H | 300 | Mass length | a | 7000 |

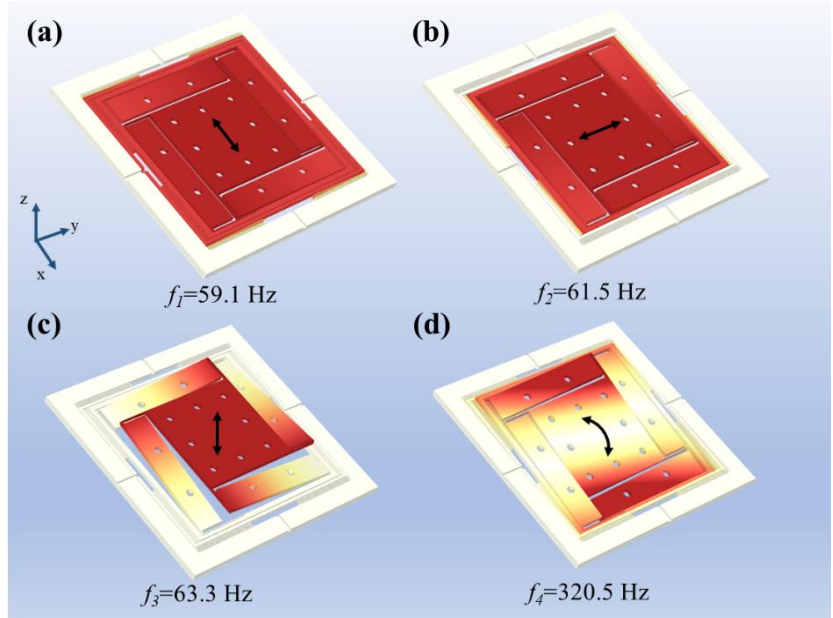

Fig.3 Simulation results of the resonant frequency of the three-axis accelerometer. (a) The first-order resonant frequency $f_1$ is 59.1 Hz, corresponding to the vibration mode along the X-axis within the plane. (b) The second-order resonant frequency $f_2$ is 61.5 Hz, corresponding to the vibration mode along the Y-axis within the plane. (c) The third-order resonant frequency $f_3$ is 63.3 Hz, corresponding to the vibration mode along the Z-axis outside the plane. (d) The fourth-order resonant frequency $f_4$ = 320.5 Hz corresponds to the rotational motion mode.

Fig. 3 presents the simulated vibration modes of the triaxially integrated accelerometer

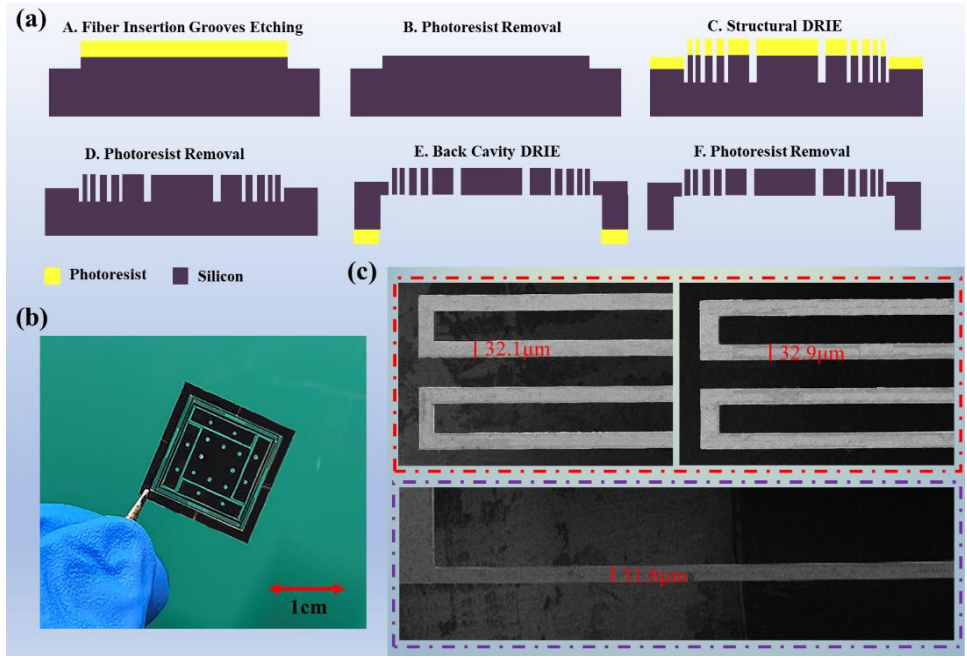

Fig. 4 (a) The fabrication process of the proposed accelerometer chip. (b) The physical picture of the fabricated sensor chip before inserting the fiber. (c) The red frames show the SEM images of the X-axis and Y-axis folding spring beams, while the purple frames depict the CD section of the U-shaped beam.

designed with parameters from Table 1. Fig.3(a) to (c) display the fundamental vibration modes along the X-axis, Y-axis, and Z-axis, with resonant frequencies of 59.1 Hz, 61.5 Hz, and 63.3 Hz, respectively. Fig. 3(d) shows a representative higher-order vibration mode at 320.5 Hz. The significant difference between the fundamental resonant frequency and the higher-order resonant frequency effectively suppresses orthogonal cross-axis interference within the sensor.

**B. Sensor Fabrication**

The proposed accelerometer was fabricated using MEMS processes, as shown in Fig.4(a). Firstly, a 500-μm-thick silicon substrate underwent plasma cleaning. And four 200-μm-deep grooves were etched into the substrate to accommodate optical fiber insertion. Subsequently, the main 300 μm deep structure was etched into the front side. Finally, a 200 μm thick layer was etched from the backside of the substrate to release the central spring-mass accelerometer structure. Between each etching step, plasma enhanced chemical vapor deposition (PECVD) was employed to spin coat 2μm photoresist on the silicon substrate and patterned using photolithography development technology to define the mask for the subsequent etch. The photoresist was stripped following each etching process. Fig. 4(b) shows the fabricated accelerometer chip, with dimensions of 16 mm × 16 mm × 0.5 mm, representing significant miniaturization compared to existing triaxial optical accelerometers. Fig. 4(c) shows a scanning electron micrograph (SEM) of the fabricated device. Beam dimensional errors are within 1 μm, indicating high-precision microfabrication that minimizes deviation from theoretical designs. Furthermore, the uniform thickness design simplified processing and improved device yield.

## 3. Sensor Performance

**A. Experimental Testing**

An optical test system was established under laboratory conditions to conduct comprehensive

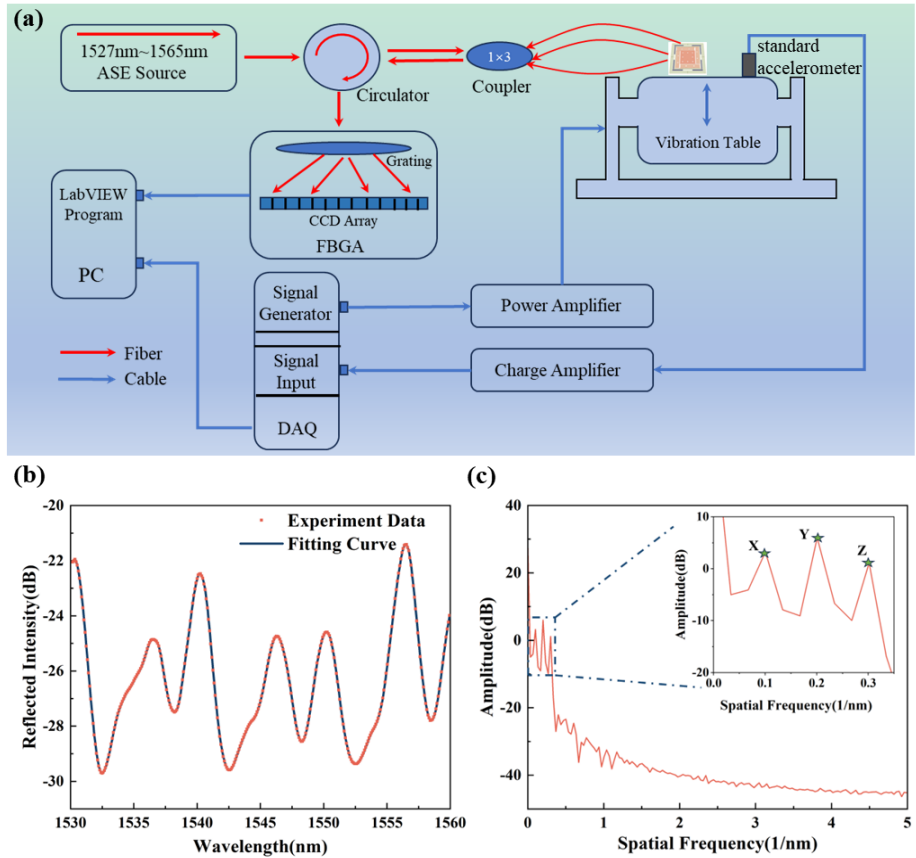

Fig. 5 (a) The schematic of the triaxial accelerometer calibration and test system. (b) The mixed interference spectrum of three FP cavities. (c) Fourier transform spectrum of the mixed interference spectrum. Points X, Y, and Z respectively correspond to the FP cavities of the three axes.

performance evaluation of the fabricated accelerometer. As shown in the Fig 5(a), A C-band amplified spontaneous emission (ASE, Beogold Technology) light source module serves as a broadband source to light up the proposed sensor with a wavelength range of 1526~1563 nm. The ASE broadband source is directed through a fiber-optic circulator into a 1×3 fiber coupler, which distributes the optical signal to the three orthogonally oriented Fabry-Pérot (FP) cavities corresponding to the X-, Y-, and Z-axes. The interference signal formed after reflection is input to a fiber bragg grating analyzer(FBGA, I-MON 512HS, Ibsen Photonics). Real-time acquisition of interferometric spectral information occurs, followed by input into a computer. Triaxial signals were separated in the spectral domain by their distinct frequency signatures, followed by independent demodulation of each axis using white-light phase demodulation techniques. This algorithm has been integrated into the LabVIEW program. Driven by the sinusoidal signal amplified by the power amplifier, the vibration table generates vibration. The standard accelerometer can obtain the amplitude and frequency of the acceleration signal through the charge amplifier, and a data acquisition card (DAQ, B&K 3160-A-022) is used for the output collection.

To precisely isolate triaxial signals while enabling high-accuracy simultaneous demodulation, the initial FP cavity lengths must be differentially designed. Given the 0.16 nm wavelength resolution of the fiber Bragg grating analyzer (FBGA), this work implements cavity length differences of 400 μm, 240 μm, and 150 μm for the three axes, respectively, corresponding to FSR

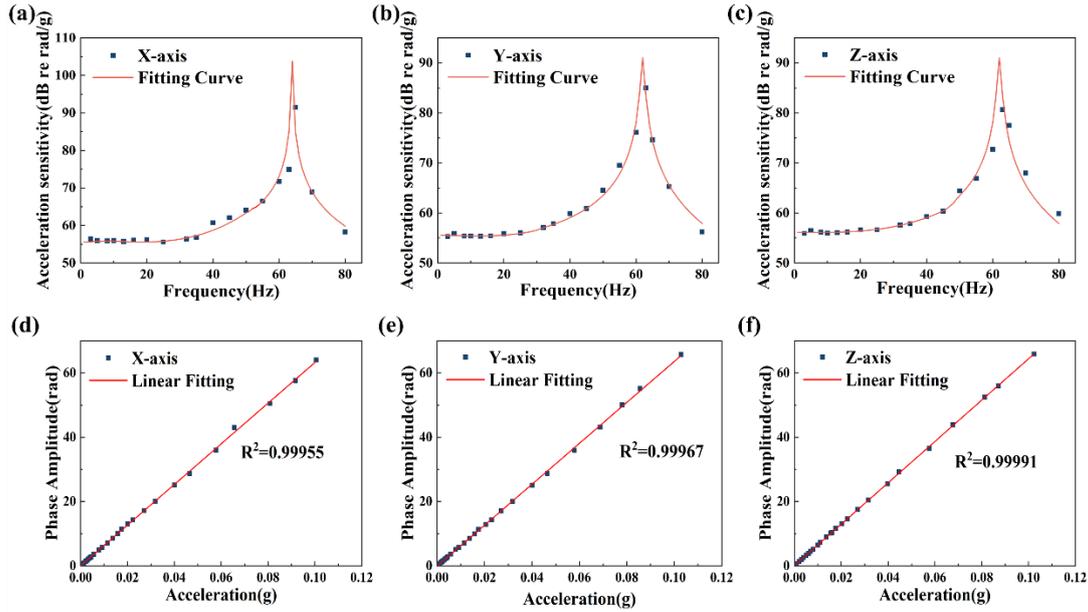

Fig. 6 (a-c) Frequency response experimental results of the X-axis, Y-axis, and Z-axis, respectively. (d-f) Linearity of the X-axis, Y-axis, and Z-axis at 10 Hz, respectively.

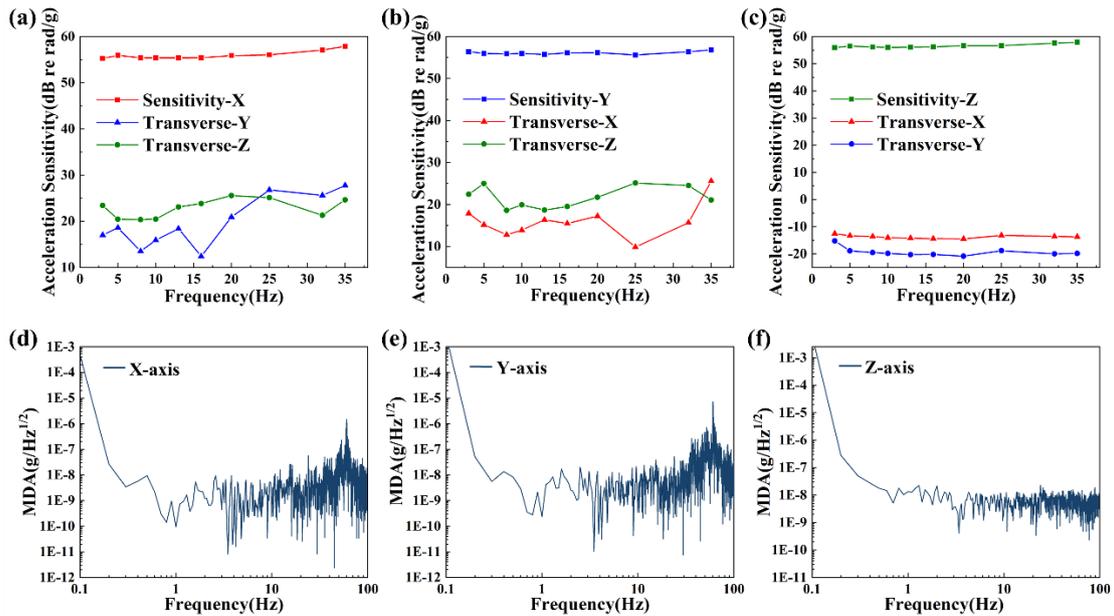

Fig. 7 (a-c) Crosstalk test results of the X-axis, Y-axis, and Z-axis, respectively. (d-f) MDA of the X-axis, Y-axis, and Z-axis at 10 Hz, respectively.

values of 3 nm, 5 nm, and 8 nm. Fig. 5(b) displays the superimposed interference spectrum of the three FP cavities, while Fig. 5(c) presents its spatial frequency distribution after fast Fourier transform processing. Three distinct spectral peaks corresponding to each sensing axis are clearly resolved, confirming effective signal isolation.

**B. Experimental Result**

    **Frequency Response and Linearity.** Fig. 6(a)-(c) presents the measured sensitivity and fitted

Table 2. Comparison between different miniature optical accelerometers

| Working principle | MDA (ng/Hz$^{1/2}$) | Cross-axis (X\Y\Z) | Size | Dimension | Ref. |
|---|---|---|---|---|---|
| FPI (MEMS) | 2.5 | \ | 10mm×10mm×0.5mm | 1 | [11] |
| Grating interferometry (MEMS) | 2 | \ | 15mm×15mm×3mm | 1 | [17] |
| MI (MEMS) | 40 | 1.53% | 21.5mm×18mm×0.38mm | 1 | [26] |
| FPI (MEMS) | 21.8 | 1.19% 1.45% 1.76% | 28mm×27mm×0.46mm | 3 | [21] |
| Optical power (MEMS) | 56200 | \ | 4mm×4mm×1.5mm | 3 | [24] |
| FPI (Metal processing) | 4500 | 2.9% | Φ20mm | 3 | [27] |
| **FPI (MEMS)** | **4.12** | **1.81% 1.44% 0.023%** | **16mm×16mm×0.5mm** | **3** | **The present work** |

curves for the X-, Y-, and Z-axes of the triaxial accelerometer. The resonant frequencies of all three axes occur near 60 Hz, consistent with simulation results. The observed 1-3 Hz frequency deviations are attributed to manufacturing error. The measured acceleration sensitivities are 55.5 dB rad/g (X-axis), 55.6 dB rad/g (Y-axis), and 56.2 dB rad/g (Z-axis), respectively. With less than 2 dB variation across axes, the device demonstrates good cross-axis consistency. It exhibits a flat response within the 1–35 Hz frequency range, with a deviation within ±1.5 dB. Due to the limitations of the laboratory vibration table, vibration signals below 1 Hz could not be detected, but they can still respond theoretically. Subsequently, the sensor's linearity and dynamic range were tested, as shown in Fig. 6(d)-(f). By keeping the fixed acceleration signal frequency at 10 Hz and increasing the vibration amplitude of the vibration table, the phase variation of the sensor was recorded. The linearity coefficients for all three axes exceed 0.999, indicating excellent linearity and the upper detection limit of the accelerometer is 0.1g.

**Crosstalk Analysis.** For cross-axis sensitivity calibration, the accelerometer was rigidly mounted on the vibration table and rotated to apply acceleration along non-sensitive directions of each axis. Fig. 7(a)-(c) displays cross-axis interference test results for XYZ axes. Principal measurements revealed average cross-axis interference of 1.53% in the Y-direction and 2.10% in the out-of-plane Z-direction for the X-axis, 0.99% in the X-direction and 1.90% in the out-of-plane Z-direction for the Y-axis, and 0.031% in the X-direction and 0.016% in the Y-direction (both in-plane) for the Z-axis. The folded spring beams exhibit moderate cross-axis mitigation, while the specially designed U-shaped cantilever structure achieves superior interference suppression. This architecture significantly reduces in-plane cross-talk for the Z-axis sensing element, providing critical insights for future low-interference out-of-plane accelerometer designs.

**Noise Level.** Low noise performance is critical for detecting weak acceleration signals in sensors. The fabricated accelerometer was mounted on a vibration-isolated platform, and output signals were acquired in a quiet environment over an extended duration. Coherent analysis was performed on the collected data to reject environmental noise interference[29]. The resulting minimum detectable acceleration (MDA) for three axes is presented in Fig. 7(d-f). Measured MDA values at the 10 Hz center frequency of 1/6-octave bands were 4.12 ng/√Hz along the X-axis, 6.68 ng/√Hz along the Y-axis, and 5.54 ng/√Hz along the Z-axis. The developed triaxial optical accelerometer achieves ultra-low noise levels, enabling effective detection of nano-g-level accelerations. It has broad application prospects in fields such as seismic wave detection, oil and gas exploration, and aerospace for weak signal detection.

Table 2 compares the performance parameters of the proposed accelerometer against the latest proposed miniature optical devices. Our single-proof-mass triaxial accelerometer achieves three-axis acceleration sensing within a miniaturized footprint while significantly mitigating orthogonal cross-axis interference in the Z-axis—a persistent challenge for monolithic designs. With an ultra-low intrinsic noise floor of 4.12 ng/√Hz, the sensor enables precise micro-vibration detection. This combination of compact integration, crosstalk suppression, and exceptional resolution demonstrates strong potential for low-frequency vector sensing applications such as seismic wave detection.

## 4. Conclusion

In summary, this work presents a monolithic multi-degree-of-freedom optical accelerometer fabricated via MEMS processes, enabling triaxial acceleration measurement with ultralow noise. Unlike conventional multi-chip solutions, the proposed design employs a single proof mass shared across all three axes, eliminating alignment errors inherent in assembled configurations while achieving size reduction. In response to the poor consistency and crosstalk performance issues of the triaxial single mass block accelerometer, this paper proposes a special U-shaped beam as the out-of-plane sensitive structure, which effectively suppresses the crosstalk of the out-of-plane vibration structure, and by designing the parameters of the in-plane and out-of-plane structures, good consistency in three axes is achieved. The optical detection method enables this sensor to have advantages such as anti-electromagnetic interference, all-optical passivity, and long-distance detection. The final experimental test shows that the size of the fabricated accelerometer chip is 16mm×16mm×0.5mm, the sensitivity difference between the three axes is less than 2 dB, and the crosstalk of the out-of-plane sensitive structure is 0.023%. Moreover, the noise level is 4.12 ng/√Hz @10 Hz. To our knowledge, it is the only multi-degree-of-freedom triaxial optical accelerometer that simultaneously achieves small size, extremely low noise and low crosstalk.

As research on monolithic triaxial optical accelerometers remains limited, this study provides valuable insights for next-generation designs. Furthermore, its compact structural design makes it easy to integrate with other sensors or devices to form a three-axis precise measurement system. The low cross-interference between axes and the noise level of 4.12 ng/√Hz give it significant advantages in the field of weak acceleration signal detection, providing new solutions for earthquake wave detection, oil and gas exploration, aerospace and other fields.